# Deterministically Charged Quantum Dots in Photonic Crystal Nanoresonators for Efficient Spin-Photon Interfaces

Konstantinos G. Lagoudakis[1], Kevin Fischer[1], Tomas Sarmiento[1], Arka Majumdar[2], Armand Rundquist[1], Jesse Lu[1], Michal Bajcsy[1] & Jelena Vučković[1]

[1]E. L. Ginzton Laboratory, Stanford University, Stanford CA 94305, USA
[2]Department of Physics, University of California, Berkeley, CA 94720, USA

E-mail: lagous@stanford.edu

**Abstract**: We demonstrate a novel method for deterministic charging of InAs quantum dots embedded in photonic crystal nanoresonators using a unique vertical *p-n-i-n* junction within the photonic crystal membrane. Charging is confirmed by the observation of Zeeman splitting for magnetic fields applied in the Voigt configuration. Spectrally resolved photoluminescence measurements are complemented by polarization resolved studies that show the precise structure of the Zeeman quadruplet. Integration of quantum dots in nanoresonators strongly enhances far-field collection efficiency and paves the way for the exploitation of enhanced spin-photon interactions for fabrication of efficient quantum nodes in a scalable solid state platform.

Systems involving single quantum emitters interacting with photons are not only interesting in the context of quantum optics and cavity quantum electrodynamics but have been shown to be excellent candidates for quantum information processing and communications. Along these lines, a requirement of utmost importance is efficient interfacing of flying and stationary qubits. Some of the most promising platforms for the realization of such interfaces, are superconducting circuits,[1],[2] , trapped ions and ion chains[3],[4], nitrogen vacancies in diamond[5],[6], and atoms in cavities[7],[8]. The platform of photonic crystals with embedded quantum dots (QDs) is another very promising system where efficient interfacing of flying and stationary qubits can be implemented for the creation of scalable quantum networks and quantum information processing[9].

Although neutral QDs could be used as stationary qubits, they suffer from relatively short coherence times[10] on the order of a nanosecond. On the other hand, the ground states of quantum dots containing an additional electron are characterized by coherence times that are many orders of magnitude longer[11] (100s of μsec) and therefore are much better suited for use as qubits. Furthermore, the optical addressability of these states allows for ultrafast control.

Probabilistically charged quantum dots in bulk materials have been successfully used for qubit initialization and manipulation[12], but the low charging efficiencies and the probabilistic nature of the charging of the QDs present serious limitations to this approach. Several methods have been proposed and implemented for deterministic charging of QDs in order to eliminate these issues: Schottky junctions[13] or lateral[14] and more commonly vertical *p-i-n* junctions[15]-[17].

Integration of junction-embedded photonic crystal nanoresonators with quantum dots allows for significant efficiency improvements. In addition to the enhancement of far-field collection efficiency, the ability to confine light to extremely small mode volumes (of the order $(\lambda/n)^3$) greatly enhances the interaction between photons and quantum dots.

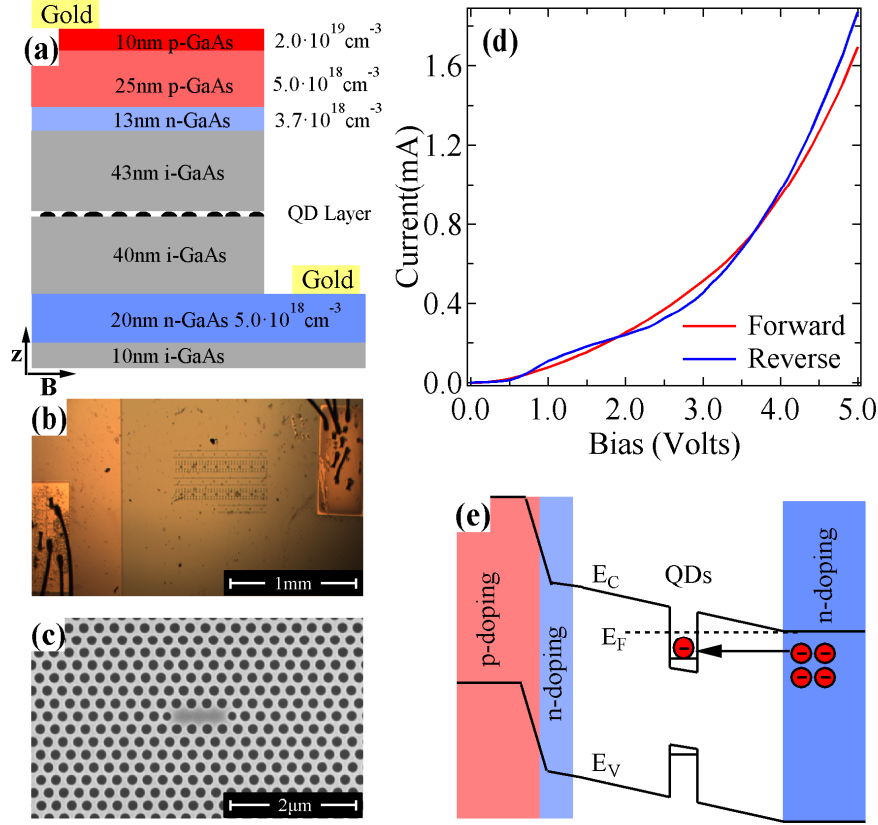

**Figure 1**: (a) Photonic crystal membrane composition and doping profile. The quantum dots are depicted as the black domes at the center of the membrane. The arrows show the main axes of the structure: z corresponds to the growth direction and B is the direction of the magnetic field (applied later on) (b) Low magnification optical microscope image of the sample. The wire bonded contact pads are seen at the two sides while an array of nanocavities can be seen between the pads. The lightly shaded region around the left pad corresponds to the etched part that exposes the bottom n-doped layer. (c) SEM image of an L3 cavity. (d) I-V curve of the sample at cryogenic temperature. (e) Schematic band diagram of the *p-n-i-n* junction. $E_V$ is the valence and $E_C$ the conduction band and $E_F$ corresponds to the Fermi level. Layer thicknesses are not up to scale.

Several geometries have been proposed for interfacing charged QD spin to photons by means of either orthogonally polarized degenerate H1 cavities[18] or linearly polarized defect L3 cavities. Very recently, experimental efforts based on nearly resonant L3 nanoresonators have shown exciting results on spin initialization and manipulation[17]. Since the metallic films required for Schottky junctions drastically reduce nanoresonator quality factor, *p-i-n* junctions are better suited for use with the photonic crystal platform. Yet, while *p-i-n* structures have been very successful for deterministic charging, several issues need to be addressed, most notably the relatively high bias involved for charging the dots and the difficulty in imposing the sign of the charging.

Here we investigate a novel structure for deterministic charging of QDs in photonic crystal nanoresonators, based on a vertical *p-n-i-n* geometry that has reduced built-in bias while allowing for negative charging of the QDs on demand. We perform a detailed magneto-spectroscopic study of QDs in close resonance to photonic crystal nanoresonators and demonstrate charging by the application of a strong magnetic field in the Voigt configuration that results in the characteristic quadruplet Zeeman splitting of charged dots[19].

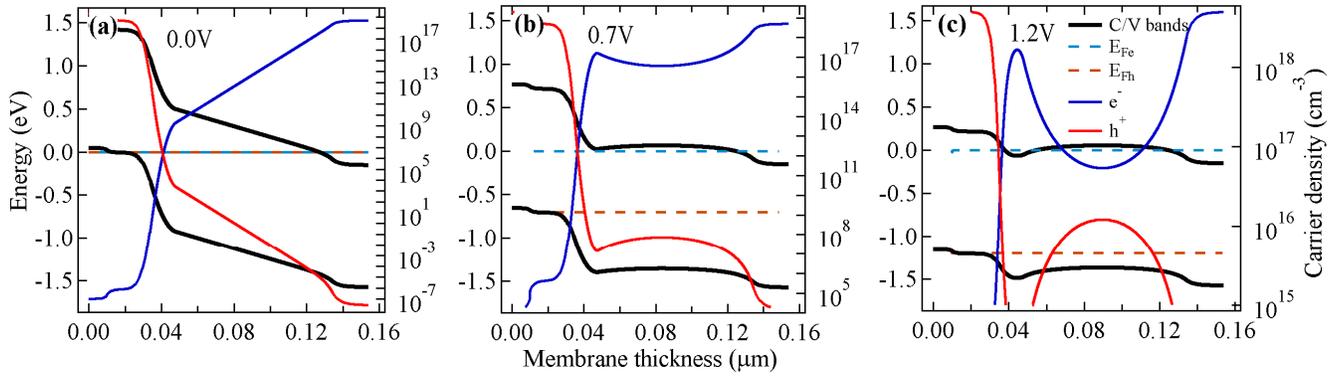

**Figure 2**: Simulated band structure (black lines), quasi-Fermi levels for electrons (orange) and for holes (cyan) and carrier densities (red for holes and blue for electrons) along the growth direction for three different forward biases (a) 0 volts, (b) 0.7 volts and (c) 1.2 volts. The carrier densities are plotted against the right-hand axis whereas the energies of conduction and valence bands and the quasi-Fermi levels are plotted against the left-hand axis. The inbuilt field that bends the bands at zero bias is counterbalanced when a forward bias is applied, resulting in a flattening of the bands. The high electron density at the central region (80nm beneath the membrane surface) where the quantum dots are situated ensures negative charging of the quantum dots. For simulations, we employ Sentaurus solver.

The sample that was used in this study consists of a 164 nm GaAs membrane with a *p-n-i-n* doping profile (figure 1(a)), containing InAs QDs in the center of the intrinsic region. An 873 nm thick $Al_{0.8}Ga_{0.2}As$ sacrificial layer leaves the membrane suspended when undercut. A distributed Bragg reflector consisting of 5 alternating GaAs-$Al_{0.8}Ga_{0.2}As$ quarter-wavelength layers was grown underneath the sacrificial layer in order to enhance collection efficiency. Gold pads were deposited on both sides of the sample, as shown in figure 1(b), in order to electrically contact the device.

Electron beam lithography followed by inductively coupled plasma etching was used to pattern the arrays of nanoresonators. A scanning electron microscope (SEM) image of an L3 cavity is shown in figure 1(c). The lower *n*-doped region was exposed using a Piranha wet etch, allowing for both the *p* and *n* layers of the junction to be contacted[9].

The sample was held at approximately 10 K on the cold finger of a continuous helium flow magnetic cryostat the superconducting coil of which can create a maximum field of $B_{max}$=5T. In order to apply a magnetic field in the Voigt configuration, the sample was oriented with its growth direction perpendicular to the magnetic field lines. A special holder with a mirror oriented at 45° above the sample allows for easy optical access of the sample surface. QD luminescence was collected with a long working distance microscope objective with NA=0.5 followed by a standard confocal microscopy setup with polarization resolution. A polarization maintaining 10:90 beam splitter was used to minimize losses of the collected luminescence. QD luminescence was generated by pumping with an above band continuous wave (cw) laser at 780 nm. The photoluminescence spectra were recorded at the output of a 0.75 m long monochromator with a ~45 μeV resolution. In order to study the junction's electrical response, an ultra-stable power supply was used to scan the applied voltage across the junction. A typical I-V curve of the sample studied here is provided in figure 1(d) showing diode-like characteristics. Its non-ideal behavior comes from residual resistive channels created through fabrication imperfections such as spurious gold deposition on the sample edges, degradation of Ohmic contacts due to oxidation, and possible gold diffusion through the membrane during wire-bonding. These resistive channels also lead to a reduced bias across the *p-n-i-n* junction with respect to the externally applied bias.

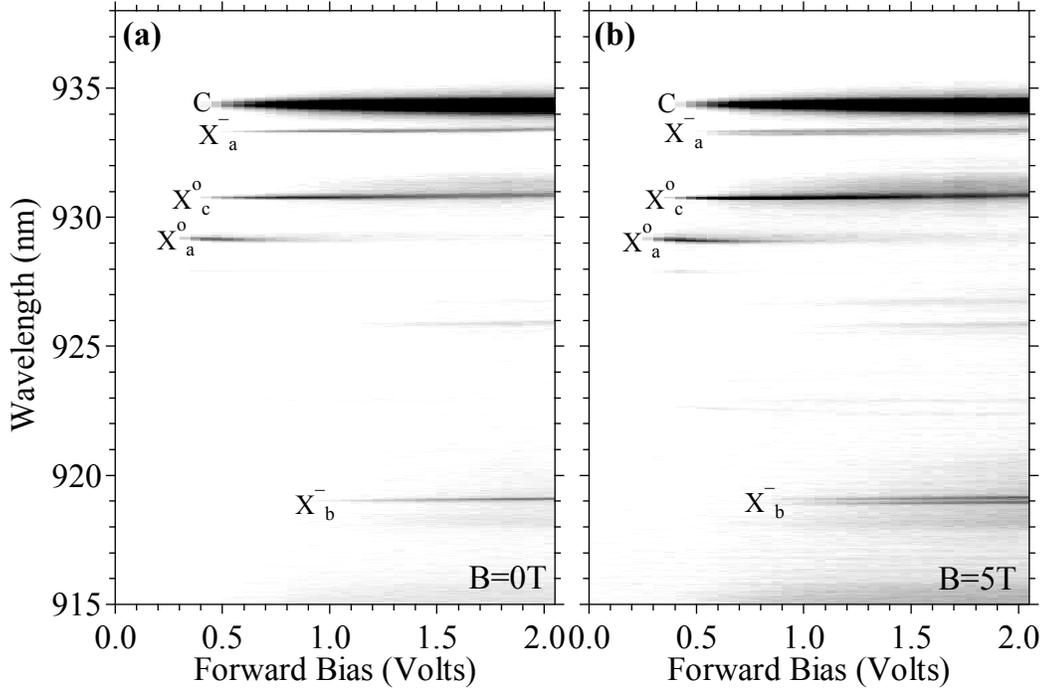

**Figure 3**: Overview of all spectral lines for B=0T (a) and B=5T (b) as a function of applied bias. The lines are labeled as: nanocavity mode C, neutral excitons $X_a^o, X_c^o$ and charged excitons $X_a^-, X_b^-$. As the QD gradually charges, the neutral $X_a^o$ line disappears while the trion $X_a^-$ line appears, and beyond 1.2V the QD is charged with maximal probability. Although spectral jumps like the one of $X_a^o$ to $X_a^-$ are good indicators of a charging event as a function of the bias, it is not enough to prove charging. This is more so in the case of charged quantum dots like $X_b^-$ which here appears at 0.9 volts bias and has no identifiable associated neutral exciton. When a magnetic field is applied in the Voigt configuration though, charged excitons can be easily identified as they become doublets for high magnetic field.

The *p-n-i-n* structure is designed to favor negative charging of the QDs by blocking hole injection at the *p-n* junction. The doping and thicknesses of these *p* and *n* layers are such that the energy bands in the intrinsic region are slightly tilted for zero forward bias, as can be seen in the schematic representation of figure 1(e). Application of a forward bias pushes the excess carriers of the bottom *n*-doped region further in the intrinsic region where the QDs are located, favoring injection of electrons into the QDs and therefore favoring negative charging. In addition, the application of a forward bias counter balances the effect of the built-in field, resulting in a flattening of the bands as shown in figure 2(b) and 2(c).

For low forward biases carriers that are generated optically by the above band excitation in the vicinity of the QDs are separated in the intrinsic region of the junction because of the inbuilt field and the wavefunctions of electrons and holes within the quantum dots are spatially mismatched. The cumulative effects of charge separation and spatial mismatch lead to weak photoluminescence intensity. As the conduction and valence bands flatten out at higher forward bias we observe an increase in the photoluminescence intensity as well as a gradual shift of the QD towards higher energies as a result of the quantum confined Stark effect (QCSE). This behavior can be seen in figure 3 that provides the spectral response of a typical device as a function of the applied bias.

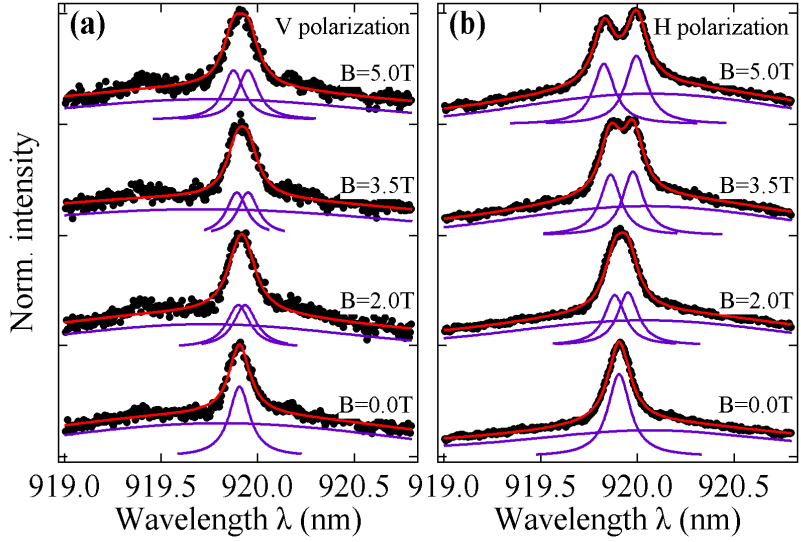

**Figure 4**: Multi-fits of the spectral response of a typical trion for four values of the applied magnetic field in the Voigt configuration for vertical (a) and horizontal (b) polarizations. Black points correspond to the raw spectra. Violet solid lines are the individual peaks that were fitted, and the sum of all violet lines gives the overall fit (red solid line). The linewidth used for the multi-fits is extracted from the linewidth of the dot for zero magnetic field.

For increasing bias the QD lines ($X_i^j$) and consequently the cavity mode (C) become gradually more visible while the neutral exciton $X_a^o$ centered at ~929nm slightly shifts to higher energy (by ~0.1nm) because of the QCSE. The presence of extra carriers in the quantum dots due to the increased bias facilitates charging of the excitons. When an extra electron binds to the exciton, the energy of the complex drops and a new spectral line corresponding to the charged exciton $X_a^-$ appears at ~933.5nm. The relative intensity ratio between $X_a^-$ and $X_a^o$ gives a good estimate of the charging probability, and as a figure of merit, here the $X_a^-$ is charged with ~95% probability for 1.2V bias.

It is worth noting that the onset of device operation is ~0.25V and the spectral lines persist for biases as high as 6V (maximum of our precision voltage supply). For forward biases beyond 2V, however, ohmic heating results in a global redshift of cavity and QD lines.

Although spectral 'jumps' like the one described above provide strong indications of charging, to conclusively demonstrate it, it is necessary to perform the same experiment under a strong magnetic field in the Voigt configuration ($\vec{B} \perp \vec{z}$ as shown in figure 1(a)). The excited states of a trion in the absence of magnetic fields contain two electrons and a hole, whereas the ground states contain a single electron. Application of a magnetic field in the Voigt configuration independently mixes the excited states as well as the ground states of the trion, which results in four optically active transitions with different energies. The exact transition energies are defined by the Zeeman splitting of the ground and excited states of the trion system. The splitting of the ground states is determined solely by the Landé g-factor of the electrons, whereas the splitting of the excited states is purely determined by the *g*-factor of holes (the electron pair share the same orbital and due to the Pauli exclusion principle is in a spinless singlet state). The selection rules that govern the quadruplet impose orthogonal linear polarization between the two outer and two inner transitions [19].

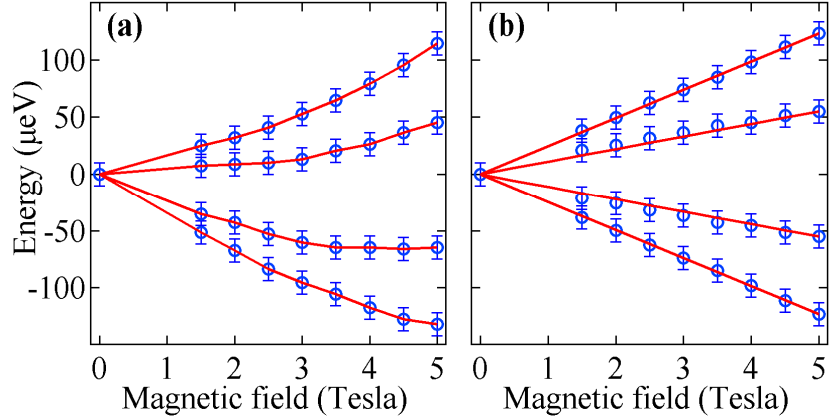

**Figure 5**: (a) Locations of the fitted peaks from the raw data. The diamagnetic shift along with any temperature fluctuations bend the splitting lines. (b) Pure Zeeman splitting after removal of the diamagnetic shift.

Upon application of the magnetic field, it is therefore expected to observe one spectral doublet in one polarization and another doublet with slightly different energies for the opposite polarization. Contrary to this characteristic behavior of charged excitons, neutral excitons do not show any splitting for the magnetic fields achievable with our superconducting magnet.

Figure 3 (b) provides the spectral landscape of figure 3 (a) but for a 5T magnetic field in the Voigt configuration, as a function of applied bias. A careful inspection of lines $X_a^-$ and $X_b^-$ reveals that they both have become doublets with applied magnetic field due to the Zeeman interaction, demonstrating that these lines correspond to trions. It is also worth noting that although the $X_a^-$ has a neutral exciton associated to it, $X_b^-$ has no identifiable associated neutral exciton, making it particularly difficult to classify as a trion before the application of the magnetic field.

To further investigate and characterize the properties of the Zeeman splitting, we performed a complete polarization analysis on $X_b^-$. We chose the charged line furthest from the nanoresonator frequency in order to avoid issues related to the polarization selectivity of the cavity. We set the forward bias to 1.2V, where the trion photoluminescence intensity is close to maximum, and we scanned the magnetic field amplitude over the whole available range (0 to 5T).

When we apply the magnetic field, the spectroscopic signature of the trion radically changes showing significant broadening at low fields that develops into a full quadruplet structure at high magnetic field. The locations of the individual peaks were extracted from the spectra by decomposing the overall spectral response into Lorentzian peaks by least squares regression. The individual fitted peaks along with the overall fit and raw data are shown in figure 4 for the two polarizations. The linewidth used for the multi-fits is that of the individual trion transitions for zero applied magnetic field, here being of the order of 0.1 nm. This width is a convolution of the actual QD lineshape with the response function of the monochromator, which is highly dependent on the monochromator slit width and grating type. The peak locations of the multi-fits for all the magnetic fields are shown in figure 5. The trend of the splitting is given by the sum of the linear Zeeman splitting, a global quadratic shift related to the diamagnetic shift and any temperature fluctuations while changing the magnetic field. The raw data peak locations are shown in figure 5(a) while the actual linear Zeeman splitting after removal of the diamagnetic shift is shown in figure 5(b).

The slope of the outer and inner transitions provide the sum and difference of the Landé g factors of electrons and holes, from which we extracted $g_e$=0.25±0.05 and $g_h$=0.60±0.01, in agreement with recent literature [20]. Contrary to what was recently reported on the dependence of the *g* factors of electrons and holes as a function of the applied bias in a *p-i-n* structure [20], here we do not observe a similar effect. The splitting and polarization properties remain the same for a wide range of applied biases.

The device presented here utilizes a novel structure for deterministic negative charging of QDs in close resonance to photonic crystal nanoresonators. Compared to the usual *p-i-n* doped membrane profile, the *p-n-i-n* profile used here, inherently flattens the bands and quantum dots become optically active at voltages as low as 0.25V. Simultaneously, it keeps the positive excess carrier density low in the vicinity of the QD, while the *n* region on the bottom of the structure provides the high electron density to allow for deterministic negative charging of the QDs. The charging is demonstrated conclusively by the application a magnetic field in the Voigt configuration and the observation of a quadruplet splitting of the trion resonance. Inversing the doping profile should allow for positive charging of the QDs on demand, which could be of future interest [21]. Future investigations will employ resonant excitation for the investigation of the charged QD linewidth and the general suitability of this device for spin photon interfaces.

**Acknowledgements**

The authors acknowledge financial support provided by the Air Force Office of Scientific Research, MURI Center for Multi-functional light-matter interfaces based on atoms and solids. K.G.L. acknowledges support from the Swiss National Science Foundation. K.F acknowledges support from the Stanford Graduate Fellowship.